# Nanoscale spectroscopic studies of two different physical origins of the tip-enhanced force: dipole and thermal

Junghoon Jahng[†], Sung Park[§], Hyuksang Kwon[†], Derek Nowak[§], William A. Morrison[§], Eric O. Potma[∥]* and Eun Seong Lee[†]*

[†]Center for Nanocharacterization, Korea Research Institute of Standards and Science, Daejeon 34113, South Korea.

[§]Molecular Vista Inc., 6840 Via Del Oro., San Jose, CA 95119, USA

[∥]Department of Chemistry, University of California, Irvine, CA 92697, USA

*Corresponding authors. Email: eslee@kriss.re.kr (E. S. L.); email: epotma@uci.edu (E. O. P.)

**ABSTRACT:** When light illuminates the junction formed between a sharp metal tip and a sample, different mechanisms can contribute to the measured photo-induced force simultaneously. Of particular interest are the instantaneous force between the induced dipoles in the tip and in the sample and the force related to thermal heating of the junction. A key difference between these two force mechanisms is their spectral behaviors. The magnitude of the thermal response follows a dissipative Lorentzian lineshape, which measures the heat exchange between light and matter, while the induced dipole response exhibits a dispersive spectrum and relates to the real part of the material polarizability. Because the two interactions are sometimes comparable in magnitude, the origin of the nanoscale chemical selectivity in the recently developed photo-induced force microscopy (PiFM) is often unclear. Here, we demonstrate theoretically and experimentally how light absorption followed by nanoscale thermal expansion generates a photo-induced force in PiFM, which hasn't been revealed so far. Furthermore, we explain how this thermal force can be distinguished from the induced dipole force by tuning the relaxation time of samples. Our analysis presented here helps the interpretation of nanoscale chemical measurements of heterogeneous materials and sheds light on the nature of light-matter coupling in van der Waals materials.

The blossoming field of nanotechnology has triggered a demand for characterization tools that enable compositional analysis with nanoscale spatial resolution. Traditional optical microscopic methods have insufficient resolution to meet this need, and even super-resolution fluorescence microscopy techniques are too limited in their applicability for the analysis of a broad range of nano-materials[1]. Scan-probe techniques, on the other hand, offer genuine nanoscale resolution but generally lack chemical contrast. To overcome these limitations, recent efforts have focused on technologies that combine the high-resolution capabilities of scan-probe techniques with the chemical selectivity of optical spectroscopy. The scattering type scanning near-field optical microscopy (s-SNOM) is a prime example, as it enables chemically sensitive imaging of materials at the nano-scale[2]. In addition, the light-based scan-probe approach also enables the study of fundamental light-matter interactions in the tip-sample junction, such as polaritonic effects in van der Waals materials[3-4].

As a nanoscopic technique, s-SNOM has made it possible to characterize physical and chemical properties of nanoscale materials with great success. Nonetheless, experimental challenges remain. For instance, it is still challenging to isolate the actual near-field response from the background, often requiring complex methods for background suppression such as the pseudo heterodyne technique[5] or the higher order demodulation technique[6]. Because of the low signal-to-noise ratio, s-SNOM typically uses moderately high power, which may cause damage to the light sensitive nano-junction and irreversibly alter the sample properties on the molecular level.

An alternative near-field method uses the opto-mechanical force as the read-out mechanism, thus avoiding the detection of light altogether. These techniques include photo-induced force microscopy (PiFM)[7-8], photothermal-induced resonance microscopy (PTIR)[9-11] and peak force infrared microscopy (PFIR)[12]. Both PTIR and PFIR probe the spectroscopic response of the sample by registering the thermal expansion of the material, which is in mechanical contact with the tip and in contact resonance mode. Although these latter approaches generate nanoscale maps with genuine chemical contrast, they require the tip to be in physical contact with the sample. In some cases, scanning in contact mode may pose problems and can damage the sample[13]. Compared to the PTIR and PFIR, the PiFM operates in the non-contact/tapping mode while monitoring the interaction force between photo-induced dipoles in the tip and in the sample. Theoretically, the spectral response of induced-dipole forces follows a dispersive lineshape implying a different image contrast. Indeed in some reported spectral PiFM measurements the dispersive lineshape is observed[9-11]. However, recent PiFM experiments in the mid-IR have produced dissipative line shapes as in the PTIR[10] and PFIR[12]. The latter observations suggest that optical absorption followed by a thermal process is somehow involved. Thus a question is raised about how thermal contributions affect the PiFM signal and whether they may overwhelm the instantaneous induced-dipole forces.

In this work we reveal how the thermal process constitutes another force generation mechanism in the PiFM measurement. Moreover, we discuss several ways to distinguish the two different mechanisms - thermal versus dipole interaction - by introducing the relaxation time as determined by the sample volume. A rigorous theoretical description in the dipole-approximation and the corresponding experimental demonstration in the illuminated tip-sample geometry will be presented.

## RESULTS AND DISCUSSION

The schematic diagrams of the induced dipole and the thermal interaction are sketched in Figure 1a and 1b respectively. For the induced-dipole interaction, the tip induces an image dipole in the sample, which then mutually interacts with the tip-dipole. A Coulombic force is generated between the induced dipoles of the tip and the sample, which is given as[7,14]:

$$F_{dip} \approx -\frac{3\text{Re}\{\alpha_t^* \alpha_s\}}{2\pi\varepsilon_0 (2z_0)^4}|E_0|^2 \quad (1)$$

where $\alpha_t$ and $\alpha_s$ are the complex effective polarizability of the tip and sample, respectively, within the point dipole approximation, $z_o$ is the distance from the center of the dipole to the surface, and $E_o$ is the incident field. In classical near-field scattering theory, $\alpha_s$ can be successfully interpreted with the help of an image dipole model, which yields $\alpha_s=\beta\alpha_t$, where $\beta$ is the complex electrostatic reflection coefficient, given as $\beta=\varepsilon-1/\varepsilon+1$. The geometrical effect of the tip such as the lightening rod effect should be considered to rigorously describe the dipole force in the tip-sample junction. That effect can be included by implementing the finite dipole model[15-16]. Thus the dipole force for the layered system with the finite dipole method is given as (see section 1 in Supporting Information for details):

$$F_{dip} \approx -\frac{1}{4\pi\varepsilon_0}\text{Re}\{\frac{\beta_{X_0}|Q_0|^2}{(z_0+X_0)^2} + \frac{\beta_{X_1}|Q_1|^2}{(z_1+X_1)^2} + \frac{(Q_1)^*\beta_{X_0}Q_0}{(z_1+X_0)^2} + \frac{(Q_0)^*\beta_{X_1}Q_1}{(z_0+X_1)^2}\}|E_0|^2 \quad (2)$$

where the $Q_0$ and $Q_1$ are the induced charges on tip, $z_0$ and $z_1$ are the position of each charges, $\beta_{X_0}$ and $\beta_{X_1}$ are the electrostatic reflection factors of each of the charges for the layered system and $X_0$ and $X_1$ are the positions of the image charges in the layered sample. The total electrostatic reflection for the layered system is described as the effective electrostatic reflection coefficient $\beta_{eff}$, given as $\beta_{eff} = \frac{\beta_{X_0}+\beta_{X_1}\eta}{1+\eta}$ where $\eta = Q_1/Q_0$. The details of the derivation can be found in Section 1 of the Supporting Information.

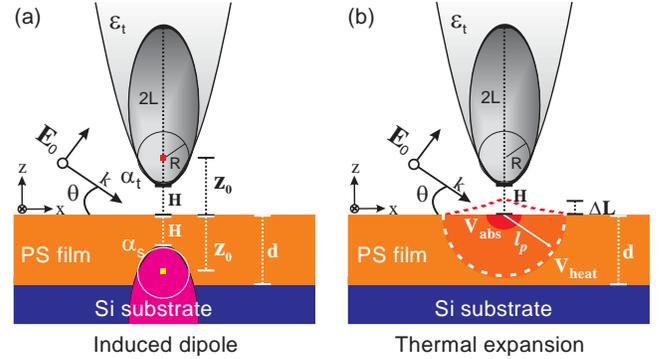

Figure 1. Schematic diagram illustrating the (a) image dipole and the (b) tip-enhanced thermal expansion. (a) $\alpha_t$ and $\alpha_s$ are the complex effective polarizabilities on the tip and the sample respectively. The tip is modeled as an ellipsoid with length of 2L. The plane wave light illuminates the sample with an angle of θ. The $z_o$ is the distance from the center of dipole to the surface. The H is the gap distance from the tip end to the sample surface. (b) The $V_{abs}$ is the absorption volume of the tip-enhanced field, the $V_{heat}$ is the heated volume related to the thermal diffusion length ($l_p$), and the ΔL is the tip-enhanced thermal expansion.

Figure 2a shows the field enhancement at the tip end with respect to the gap distance (H), for the case of a 60 nm polystyrene (PS) film on a Si substrate. The black solid line is the field enhancement at the sample with respect to the PS film thickness where H is zero (in contact). The dashed lines show the field enhancement with respect to H where the thickness of the PS is zero (blue) or 60 nm (red). The red curve is shifted by 60 nm to align with the result for H=0 nm (black solid line). Multiple-scattering effects between the tip and the substrate boosts the field compared to the case without a substrate – where the latter only includes the lightening rod effect of the ellipsoidal tip (green dashed line), given by Eq. (S1). The strong field enhancement at the Si substrate (blue dashed line) decreases as the 60 nm PS film covers the Si substrate (red dashed line). This is because the effective electrostatic reflection factor, Re{$\beta_{eff}$}, decreases from the Re{$\beta_{Si}$} (0.84) to the Re{$\beta_{PS}$} (0.43) as the PS film thickness increases, which is plotted in Figure 2b. This means that the tip becomes to recognize the layered system as sole polystyrene as the PS thickness increases. The induced dipole force by Eq. (2) is plotted in Figure 2c. The black solid line in Figure 2c is the calculated force on the ellipsoidal tip as a function of the PS film thickness where H is zero (in contact). The dash lines are the force approach curves where the thickness of the PS is zero (blue) or 60 nm (red). The red approach curve is shifted by 60 nm to align with the H = 0 nm result (black solid line). Note that, because Re{$\beta_{Si}$} is bigger than Re{$\beta_{PS}$}, the force on Si is stronger than the force on PS. This implies that if the PS clusters are prepared on the Si substrate, the image contrast is expected to be negative due to the stronger induced dipole force on Si, even at PS's absorption resonance.

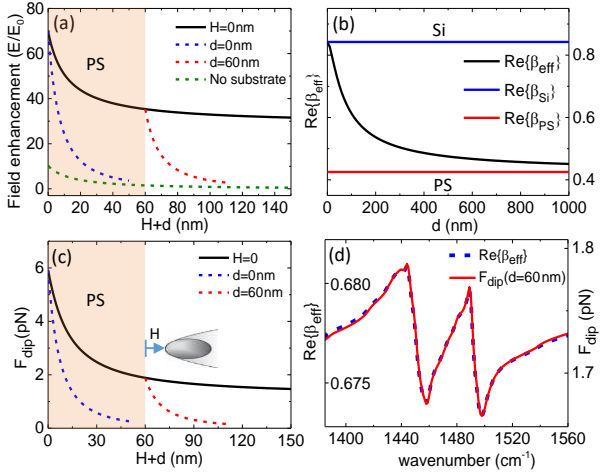

Figure 2. (a) Field-enhancement based on the finite dipole method. (b) Real part of the complex effective electrostatic reflectance factor, Re{$\beta_{eff}$}, with respect to the thickness. (c) Induced (image) dipole force based on the finite dipole method by using Eq. (2). (d) Spectrum of the real part of $\beta_{eff}$ (blue dash line) and induced dipole force (red solid line) for 60 nm PS film on the Si substrate by using Eq. (2). All the curves are calculated at the 1452 cm$^{-1}$ vibrational resonance of PS (C-H bending mode)[17].

The point dipole model is still valid to qualitatively understand the spectral response of the system. The spectral component of the force then reads Re{$\alpha_t^* \alpha_s$} ≈ Re{$\beta_{eff}$}|$\alpha_t$|$^2$ for the layered system, where the magnitude of the tip polarizability is a slowly varying function near the resonance of the sample. The real part of the effective electrostatic reflection coefficient shows a dispersive line shape resembling the index of the refraction in Figure 2d (blue dashed line). Consequently, the force spectrum shows a dispersive line shape as pointed out in previous work[8, 18-20]. The dipole force is quantitatively calculated by using Eq. (2) and plotted as a function of wavenumber in Figure 2d (red solid line). The force at the molecular absorption (weak oscillator) typically ranges from a few tens of fN to a few tens of pN[18-19].

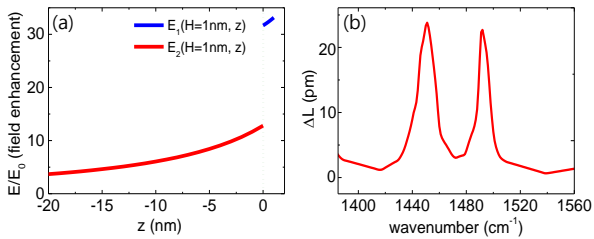

Figure 3. (a) Tip-enhanced field distribution inside (red) and outside (blue) the 60 nm polystyrene film on the Si substrate. The polystyrene interfaces with air at z = 0 nm. Left is towards the inside and right towards the outside. The tip end is located at z = 1 nm by keeping the H = 1 nm. (b) Calculated tip-enhanced thermal expansion with respect to the wavenumber for 60 nm PS film on Si substrate.

Meanwhile, near molecular resonances, there is a temperature rise of the sample due to light absorption, which results in the strain deformation of the sample that eventually gives rise to the thermal expansion. The strong field-enhancement not only enhances the induced dipole moment, but also increases the absorption of the sample[9, 11]. The diagram for the thermal response to incident light absorption is drawn in Figure 1b. According to Dazzi et al.[21], the temperature rise inside the sample is described by the heat equation with a heat source of $Q(t)$:

$$\rho C \frac{dT}{dt} = \frac{Q(t)}{V} - \kappa_{eff}\nabla^2 T \quad (3)$$

where $\rho$, $C$, and $\kappa_{eff}$ are the density, the heat capacity and the effective thermal conductivity. A rigorous modeling of the photothermal response requires a complete, multidimensional solution of the diffusion equation to be solved for the temperature distribution inside the sample which contacts the substrate acting as a thermal reservoir[22-23]. But we can gain considerable insight into the relationship between the photothermal response and the thermal properties of samples by making a simple estimate of z-directional photothermal response. The heat source $Q(t)$ is the absorbed light energy in the sample which is directly linked to the laser irradiation, given as $Q(t) = \int P_{abs}(t)\,dt$. The $V$ is the heated volume related with the absorption volume and the thermal diffusion length[24]. In this study, we consider a molecular vibration which is weakly absorptive (Im[$n$]<<Re[$n$]) where $n$ is the complex index of refraction. The sample size is much smaller than the wavelength, allowing us to use the electric dipole approximation. The absorbed power under these conditions can be expressed by $P_{abs} = \int a_{abs} \frac{1}{2} c\varepsilon_0 |E|^2 \, dV$ where $a_{abs} = \frac{4\pi}{\lambda} \frac{9Re[n]Im[n]}{(Re[n]^2+2)^2}$ which is the absorption coefficient of the sample, $\lambda$, $c$ and $E$ are the wavelength, the speed of light and the tip-enhanced electric field inside the sample respectively[25]. The above relation shows that the absorbed power $P_{abs}$ is proportional to Im[$n$]/$\lambda$, i.e., to the usual absorbance considered in infrared spectroscopy. The tip-enhanced absorption increases the temperature inside the sample and then results in the strain deformation which gives rise to the thermal expansion. One can obtain the complete time dependent temperature behavior in Ref. [26] which solves the Eq. (3). However, if we consider the maximum thermal expansion that maximally modulates the probe, the solution of the equation can be much simplified by regarding the maximum temperature change. In this case, the thermal expansion is described as:

$$\Delta L_{max} \approx \sigma d \Delta T_{max} \quad (4)$$

where $\sigma$ is the linear thermal expansion coefficient and $\Delta T_{max}$ is given as $\Delta T_{max} \approx \frac{P_{abs}}{\rho CV}\tau_{rel}$ for $\tau_{rel} < \tau_p$ and $\Delta T_{max} \approx \frac{P_{abs}}{\rho CV}\tau_p$ for $\tau_{rel} > \tau_p$[21]. Here $\tau_p$ is laser pulse width, and the $\tau_{rel}$ is thermal relaxation time which is the time for the sample to have an equilibrium with the environment as determined by the thermal diffusion process, given as $\tau_{rel} \approx 1.13\frac{4}{\pi^2}d^2/D$[26] where the $d$ is the thickness of the thin film sample and $D$ is the thermal diffusivity, given as $D = \frac{\kappa_{eff}}{\rho C}$.

The tip-enhanced field distribution inside (red) and outside (blue) of 60 nm polystyrene film on the Si substrate is calculated in Figure 3a. The polystyrene interfaces the air medium at z=0 nm. The tip end is located at z=1 nm by keeping H=1 nm. $E_1$ (blue) is the gap field between the tip and the sample and $E_2$ (red) is the field inside the 60 nm polystyrene sample. By integrating the field inside the PS and substituting it into Eq. (4), the spectral response of the tip-enhanced thermal expansion of the 60 nm PS film is plotted as a function of wavenumber in Figure 3b. The maximum thermal expansion is around 23 pm at the 1452 cm$^{-1}$ absorption resonance. The calculation details can be found in Section 2 of the Supporting Information.

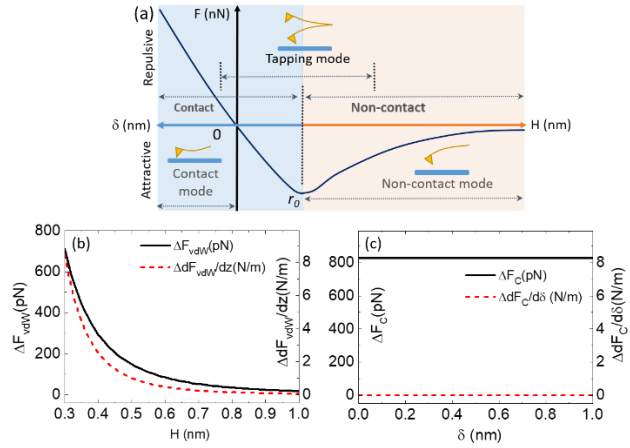

Figure 4. Attractive and repulsive tip-enhanced thermal expansion forces. (a) Force-distance diagram of AFM imaging mode with respect to the tip-sample gap distance. (b) Calculated van der Waals force change (black solid line) and the force gradient change (red dash line) due to oscillatory thermal expansion with respect to the tip-sample gap distance. (c) The calculated contact force change (black solid line) and the force gradient change (red dash line) due to oscillatory thermal expansion with respect to the indentation depth. All the calculations are obtained for the 60 nm PS on Si substrate at the vibrational resonance of 1452 cm$^{-1}$ for $\tau_p$=30 ns.

To measure the tip-enhanced thermal expansion in non-contact/tapping mode PiFM, we first have to carefully analyze the PiFM operation. The general tip-sample interaction force that probes the topography is understood as the sum of a long range attractive force and a short range repulsive force as in atomic force microscopy[14]. There are numerous models to describe the forces, especially for the contact force. In this study, we assume that the exemplar attractive force is the van der Waals force and the exemplar repulsive force is the Hookian contact force, given as:

$$F_{vdW} \approx -\frac{H_{eff}R}{12}\frac{1}{H^2} \quad (H>r_o) \quad (5)$$

$$F_c \approx k_0\delta - \frac{H_{eff}R}{12}\frac{1}{r_o^2} \quad (H<r_o) \quad (6)$$

where the $H_{eff}$ is the effective Hamaker constant between tip and sample, R is the tip radius, $r_o$ is the interatomic distance (~0.3 nm)[27], $\delta$ is the deflection of cantilever and $k_o$ is the static stiffness of the cantilever. Although both of the forces are simple descriptions of the force in the tip-sample junction, they are very helpful for understanding the typical force behavior in photo-induced force microscopy. When the force gradient is positive, there is only an attractive force, meaning that the tip does not contact the surface. On the other hand, when the force gradient is negative, a repulsive force can be exerted onto the tip, meaning that the tip might hit the surface. The positive force gradient region is called the non-contact region, while the negative force gradient region is called the contact region. In Figure 4a the non-contact region is indicated by the orange shaded area and the contact region is the blue shaded area.

PiFM is operated in the tapping mode, which covers the non-contact region and the contact region, contrary to PTIR which operates in the hard contact region (F > 0) under the contact resonance conditions[9]. It is therefore questionable how the thermal expansion of the sample can be coupled to the tip without actual mechanical 'touching', especially in the non-contact region for PiFM. A possible mechanism is explained in Ref. [28], where the sample stage was mechanically modulated with frequency $\omega_m$ and the tip was driven at $\omega_1$. When the tip approached the substrate, the heterodyne coupling frequencies $\omega_1 - \omega_m$ or $\omega_1 + \omega_m$, also known as sideband coupling, were generated in the non-contact region via the modulated van der Waals force gradient. Since the oscillating sample stage contributes to the modulation of the tip-sample gap distance, a modulation of the van der Waals force and the force gradient can be expected at $\omega_m$. In a similar manner, the oscillatory sample expansion at $f_m$ in PiFM introduces a modulation of the tip-sample gap distance in the non-contact region and eventually modulates the van der Waals force and the force gradient at $f_m$. In the contact region the modulated Hookian contact force due to the oscillatory thermal expansion pushes up the cantilever, as in PTIR operation. The modulation of the van der Waals and Hookian contact force and their force gradient due to the oscillatory sample expansion are given as below:

$$\Delta F_{vdW} \approx -\frac{H_{eff}R}{6}\frac{1}{H^3}\Delta L(H) \quad (H>r_o) \quad (7)$$

$$\Delta F_c \approx k_0\Delta L(r_0) \quad (H<r_o) \quad (8)$$

$$\Delta \frac{\partial F_{vdW}}{\partial z} \approx -\frac{H_{eff}R}{6}\frac{1}{H^3}\frac{\partial(\Delta L)}{\partial H} \quad (H>r_o) \quad (9)$$

$$\Delta \frac{\partial F_c}{\partial z} \approx 0 \quad (H<r_o). \quad (10)$$

The derivation details can be found in the Section 3 of Supporting Information. Both of the modulated forces in Eq. (7) and (8) are typically in the range between a few tens of pN to a few nN when ΔL is a few tens of pm. The force change is plotted in Figure 4b and 4c as the black solid line. Both of the forces are still 1-2 orders stronger than the induced dipole force (a few hundreds of fN to a few tens of pN). However, the modulated force gradient shows a different behavior. In Figure 4b, the modulated van der Waals force gradient, $\Delta\frac{\partial F_{vdW}}{\partial z}$, in Eq. (9) is plotted as a function of the tip-sample gap distance as the red dash line. Compared to $\Delta\frac{\partial F_{vdW}}{\partial z}$, the modulation of the contact force gradient, $\Delta\frac{\partial F_c}{\partial z}$, in Eq. (10) approaches zero which is plotted in Figure 4c as the red dashed line, because the tip-enhanced thermal expansion is maximized for the tip in contact with the surface and it doesn't depend on the cantilever deflection $z (= -\delta)$ after contact. By setting the difference frequency (or sum frequency) to match another mechanical eigenmode of the cantilever ($f_1$) as in Figure 5a, the modulated van der Waals force gradients are measured at $f_1$ simultaneously with the topography at $f_2$. The benefit of this heterodyne technique is to reject the constant background force such as scattering force and to increase the sensitivity by measuring the force gradient[7, 29-30]. The PiFM sideband mode signal is described as:

$$A_1(\omega) = \frac{k_{mod}/2}{\sqrt{m^2(\omega_1'^2 - \omega^2)^2 + (b_1'\omega)^2}} A_2 \quad (12)$$

where $k_{mod}$ is the modulated force gradient, $m$ is the mass of the cantilever, $b'$ is the effective damping of the cantilever, $A_2$ is the carrier amplitude and $\omega_1' = \sqrt{(k_1 - \frac{\partial F_{tot}}{\partial z})/m}$ [29-30]. Experimentally the non-contact region can be distinguished from the contact region in the tapping mode by monitoring the shift of the detection eigenmode of the cantilever with respect to the total applied force gradient. When an attractive (or repulsive) force affects the tip, the resonance curve is shifted to the left (or right) side because of the positive (or negative) force gradient. The total force gradient contains all the interaction force gradient of the probe which is given as $\frac{\partial F_{tot}}{\partial z} = \frac{\partial (F_{vdW}+\Delta F_{vdW})}{\partial z} + \frac{\partial (F_c+\Delta F_c)}{\partial z} + \frac{\partial (F_{dip}+\Delta F_{dip})}{\partial z}$. Because two dipole-related terms and two perturbative terms out of six terms are comparatively small, the frequency shift of the sideband amplitude peak dominantly depends on the mechanical interaction ($\frac{\partial F_{vdW}}{\partial z}$ and $\frac{\partial F_c}{\partial z}$). On the other hand, the $k_{mod}$ is the modulated force gradient, $k_{mod} = \Delta\frac{\partial F_{vdW}}{\partial z} + \Delta\frac{\partial F_c}{\partial z} + \Delta\frac{\partial F_{dip}}{\partial z}$. Because the second and the third terms are negligible, the sideband amplitude directly follows the modulated van der Waals force gradient that results from the oscillatory thermal expansion. Therefore, $\Delta\frac{\partial F_{vdW}}{\partial z}$ generates the PiFM sideband signal in both of the attractive and repulsive regions. The PiFM sideband frequency sweep in Eq. (12) is obtained by the heterodyne coupling method with the carrier amplitude under 10 nm at the set-point 88% (red) and 40% (blue) for 60 nm PS on Si substrate at 1452 cm$^{-1}$ as shown in Figure 5b. The black solid line is the measured free space resonance curve of the fundamental eigenmode by using a mechanical dither piezo. The PiFM sideband resonance curve obtained at the set-point 88% (red) is shifted to the left compared to the free space resonance curve (black). This means that the measured PiFM signal is obtained in the non-contact region (total positive force gradient). When one taps the surface, on the other hand, a repulsive force (total negative force gradient) is applied to the tip, and the sideband curves are shifted to the right side. The PiFM sideband resonance curve at the set-point 40% (blue) shows clearly the right shift. One can also see the broadened resonance curve because of the decreased quality factor resulting from the energy dissipation of the probe in mechanical contact to the sample. The PiFM spectra at the set point 88% (red) and 40% (blue) show the same dissipative curve in Figure 5c. For the set-point at 40% the signal is decreased because of the decreased sensitivity resulting from the reduced quality factor and the reduced tip-enhancement due to the tip damage in mechanical contact to the sample.

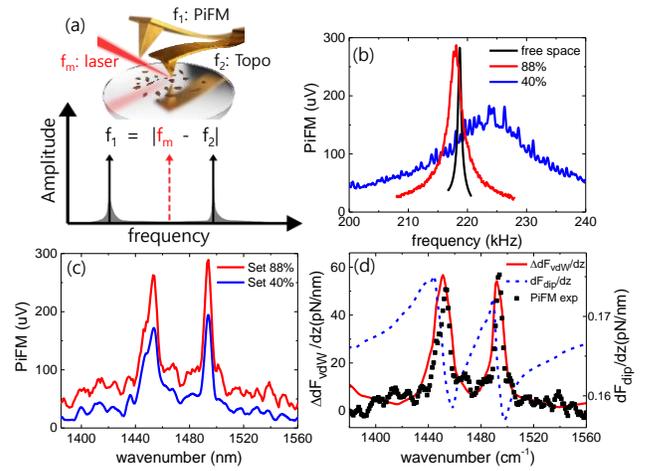

Figure 5. (a) Schematic diagram of PiFM sideband operation. (b) Measured detection resonance curves with respect to the PS film thickness at the setpoint 88% (red) and 40% (blue) at the PS vibrational resonance of 1452 cm$^{-1}$. The black solid line is the free space resonance curve. (c) PiFM spectra on 60 nm PS film at the setpoint 88% (red) and 40% (blue) at the peaks of each detection frequency in (b). (d) Calculation of the modulated vdW force gradient (red solid line) and the calculated dipole force gradient along with the PiFM data (black square dots) at the setpoint 88% for 60 nm PS on Si substrate. All the measurements are obtained under 5 mW power at a 20 μm focal spot with a 10 nm carrier amplitude.

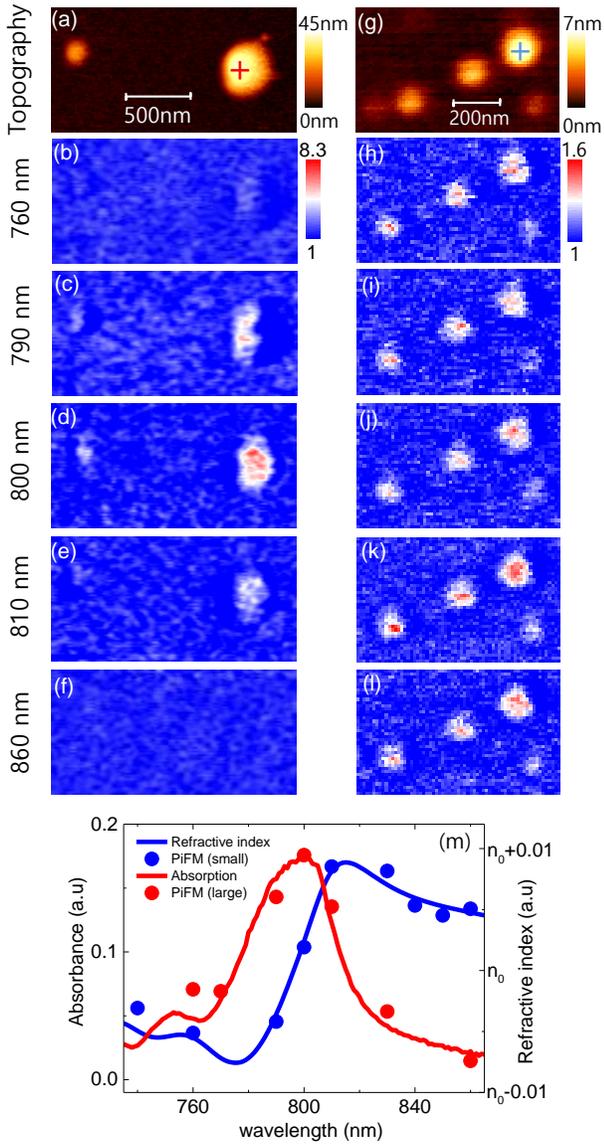

Figure 6. Chemical images of silicon 2,3-naphthalocyanine bis (trihexylsilyloxide) (SiNc) obtained with respect to the pulse width. (b) – (f) PiFM images of SiNc, obtained with CW laser, as a function of the center wavelength from 760 nm to 860 nm as well as the topography (a). (g) – (l) PiFM image of SiNc, obtained with a 200 fs laser pulse, as a function of the center wavelength from 760 nm to 860 nm as well as the topography (g). All the images are normalized by the glass substrate signal. (m) PiFM spectra (red and blue dots) obtained at '+' points on the topographies as well as the bulk SiNc spectra (red and blue solid lines). The red solid line is the absorption measurement of bulk SiNc by using UV-Vis. The blue solid line is the relative index of refraction resulting from the Kramer-Kronig relation of the absorption spectrum. $n_o$ is the refractive index at the resonance frequency. All the PiFM measurement is obtained at the setpoint 88%.

The crucial difference between the induced dipole force and the thermal force mediated by van der Waals interaction is their force gradient spectra with respect to the optical frequency. In Figure 5d, the calculated force gradient spectra of $\frac{\partial F_{dip}}{\partial z}$ (blue dash line) and $\Delta \frac{\partial F_{vdW}}{\partial z}$ (red solid line), which are based on literature values of the optical constant of polystyrene[17], are plotted along with the PiFM experimental results for the 60 nm PS film on a Si substrate at the set point 88% of 10 nm carrier amplitude. Rather than mimicking the dipole force gradient, the PiFM data corresponds remarkably well to the calculated modulated vdW force gradient. The $\frac{\partial F_{dip}}{\partial z}$ shows relatively small contrast below a few hundred fN/nm and the spectral shape is dispersive. On the other hand, the $\Delta \frac{\partial F_{vdW}}{\partial z}$ shows relatively large contrast with a few tens of pN/nm and the spectrum resembles a dissipative Lorentzian profile, which seamlessly overlaps with the measured PiFM spectrum. The calculation details can be found in Section 3 of the Supporting Information.

In Figure 6, organic dye clusters (Silicon 2,3-naphthalocyanine bis (SiNc)) whose thermal diffusivity is around 0.1 cm²/s[31] are imaged with respect to the two different force mechanisms in the near-IR range (760 nm to 860 nm). In this work, we used the tuning-fork based PiFM, which has the 32kHz fundamental resonance and the 200kHz second resonance. The second resonance probed the topography and the fundamental resonance was used for the PiFM. The left column shows PiFM images of a relatively large cluster of SiNc. In these measurements, a continuous wave (CW) laser, with an illumination power is around 200 μW and a 400 nm² focal spot, is used to demonstrate the thermal expansion behavior. The illumination power was around 200 μW at the 400 nm² focal spot. Because the laser is modulated by accusto-optical-modulator at $f_m=|f_2 - f_1|= 168$ kHz, the heating time of the sample is $1/2f_m \sim 3$ μs which is much longer than the relaxation time of the typical thin sample (< 1 μs). The right cluster in Figure 6a, which has a height of around 45 nm and a width of ~ 500 nm, thus a relatively large volume with $\tau_{rel} \sim 110$ ps, shows a strong photo-induced force signal at the 800 nm absorption resonance in Figure 6d. On the other hand, the left cluster, a height of around 25 nm and a width of ~ 200 nm, thus a smaller volume with $\tau_{rel} \sim 34$ ps, lacks clear contrast as a function of laser wavelength. It shows barely distinguishable contrast at the 800 nm absorption resonance. From the experiment we infer that the thermal expansion is not well manifested under a relaxation time of 10 ps. This volume dependence is in full agreement with our tip-enhanced thermal expansion analysis.

The spectral dependent PiFM signal reflects the dissipative nature of the thermal interaction in Figure 6b to 6f. The SiNc cluster images show a maximum contrast at 800 nm, the absorption resonance of SiNc, but almost disappear at 860 nm. The image at 760 nm shows barely discernable contrast as expected from the shape of the absorption curve of bulk SiNc by UV-Vis measurement (red solid line) in Figure 6m. The photo-induced force data (red dots) measured at the SiNc cluster, designated by a red cross '+' on the topography image in Figure 6a, follow the dissipative curve (red solid line) of the bulk SiNc spectrum as shown in Figure 6m.

The induced dipole interaction, which is independent of the relaxation time, can be separated from the thermal expansion force. It becomes dominant over the thermal effect which can be suppressed by making the relaxation time very small (under 10 ps) with a small sample volume and an ultrafast short pulsed illumination to decrease the heating time[32]. A series of the PiFM images in the right column of Figure 6 was obtained for the small size SiNc clusters by using the 200 fs pulsed laser with a repetition rate of 76 MHz. The height of the SiNc clusters in Figure 6g is under 7 nm and the width is less than 200 nm, resulting in an estimated relaxation time under 3 $ps$. Because the sample relaxation rate (~3 $ps$) is shorter the interpulse time separation (~1.32 $ns$), there is insufficient time to achieve a significant thermal expansion[33]. With the thermal contribution suppressed, the fs PiFM experiment shows a dispersive spectral response in Figure 6h to 6l. Note that we increased the incident power by 4 times (800 μW at 400 nm² focal spot) to clearly visualize the dipole contrast. The PiFM images clearly map the SiNc clusters showing a maximum contrast at 810 nm and a higher contrast at longer wavelengths (810 nm, 860 nm) than at shorter wavelengths (760 nm, 790 nm, 800 nm). Note that this behavior is very different from the longer relaxation time case (left column). The PiFM data (blue dots), measured at the SiNc cluster designated by the blue cross '+' on the topography image in Figure 6g, corroborates the dispersive curve (blue solid line) of the bulk SiNc spectrum as shown in Figure 6m. The blue solid line is the relative index of refraction resulting from the Kramer-Kronig relation of the UV-Vis absorption spectrum[34]. A fairly distinguishable contrast persists for all of the images in this spectral range. This is because the induced dipole force is based on the index of refraction, which remains finite in this frequency range. As can be seen from the dipole force and force gradient analysis for organic molecules in Figure 2d and 5d, the signal level (SiNc/glass) is much smaller than the thermal expansion dominant case and the spectral contrast is very weak because the change in the index of refraction is relatively small around the molecular resonance.

## CONCLUSIONS

The thermal expansion and the induced dipole interaction are simultaneously manifested in the tip-sample junction but they have their own characteristic behaviors. The thermal expansion force is the result of several causal processes: First, there is an energy exchange with the light field, which scales with the optical absorption coefficient and results in a temperature rise ($\Delta T \sim P_{abs}$). Second, the accumulated heat diffuses to deform the sample to induce a thermal expansion ($\Delta L \sim \Delta T$). Third, the thermal expansion changes the tip-sample distance, which introduces a modulation of the van der Waals/contact force ($\Delta F \sim \Delta L$). Because the force and the force gradient are based on the absorption, their spectrum follows the dissipative line shape. On the other hand, for the induced dipole interaction, the tip induces image charges in the sample, which mutually interacts with the tip in the near-field. The spectral dependence of the force and the force gradient follow the index of refraction so that their spectrum exhibit a dispersive line shape.

Because the induced dipole and the photo-thermal expansion depend on the material's properties and volume, we compare the typical force ranges with respect to material's characteristics in Figure 7. Even though the PiFM measures the modulated force gradient, a comparison based on force is more intuitive for comparing the two physical mechanisms. For the photo-thermal force, because the linear thermal expansion coefficient (σ) is related to the covalent bonding strength of molecules, materials with stronger chemical bonds exhibit a smaller thermal expansion. The σ of organic materials is typically in the 10-1000×10⁻⁶ K⁻¹ range, which is one to two orders of magnitude larger than the one for inorganic materials (typically 0.1-100×10⁻⁶ K⁻¹). In addition to that, the thermal diffusivity, $D \approx \frac{\kappa_{eff}}{\rho C}$, is one of the dominant factors for determining the temperature rise of the sample, which is inversely proportional to the relaxation time as $\tau_{rel} \approx 1.13 \frac{4}{\pi^2} d^2/D$. Because the thermal diffusivity of inorganic materials is around one to two orders of magnitude larger than that of organic materials[31], the thermal expansion is typically stronger in organic materials than in inorganic materials.

The dipole force for a weak oscillator such as a molecular vibrational resonance is in the range between a few tens of fN to a few tens of pN. Because our noise level is around 0.1 pN[35], the dipole force for the organic material is sometimes not measurable. However, the force is greatly enhanced for a strongly responsive oscillator such as a metal/inorganic polar crystal at plasmon/phonon polariton resonance where the permittivity is negative. The force easily reaches a few tens of pN to a few nN for the ellipsoidal tip. For these plasmonic materials, because of the large thermal diffusivity of the metal or inorganic material, the dipole force is easily extracted by suppressing the thermal force. For example, a gold nanorod shows a strong field enhancement at the tips of the structure in the polarization direction because of the localized surface plasmon. The high field distribution cannot be explained by the thermal expansion mechanism because the heat inside the gold nanorod comes from the current whose distribution peaks at the positions of the field minima (i.e. at the center) in optical antennas[36]. Thus, distinct spatial maps are expected to distinguish the two different mechanisms. Tumkur *et al.* successfully demonstrated that the dominant force for the plasmonic structure is the induced dipole (field gradient) force by mapping the field distribution of the gold nanorod whose edge is brighter and center is dark as expected[37].

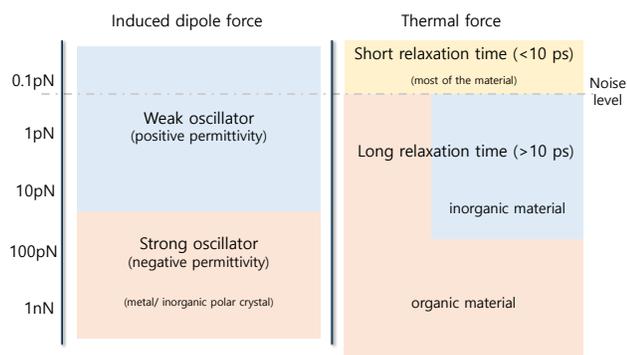

Figure 7. Typical induced dipole and thermal force ranges with respect to material's properties.

We have shown, both theoretically and experimentally that, besides the induced dipole force (pure electromagnetic effect), the tip-enhanced light absorption followed by heating and thermal expansion of nanoscale materials can also generate a photo-induced force in PiFM. The modulated van der Waals force gradient of thermally expanded samples is found to play a crucial role in exerting a net force gradient on the tip, even in the non-contact mode operation of PiFM. There is a critical distinctive characteristic of the thermal interaction, which is different from those of the induced dipole force. The difference is the spectral dependence of the signal near the resonance wavelength of the sample material. While the induced dipole interaction exhibits a dispersive spectral response, the thermal effect follows dissipative Lorentzian spectral line shape. In the view of the "chemical" imaging Lorentzian lineshapes significantly facilitate the interpretation of the molecular response, and in this regard the photothermal response may be preferred over the induced dipole force. We have suggested a way to distinguish the thermal effect from the induced dipole force by adjusting the relaxation time, which is related to the sample volume and thermal diffusivity. By measuring hyperspectral images of the photo-induced force of SiNc clusters at two different relaxation time conditions, we confirmed the different spectral dependence of the two mechanisms. The quantitative comparison between the two tip-enhanced forces helps to unravel the entwined mechanisms that give rise to photo-induced forces in hyperspectral nanoscopy. Finally, the insights obtained from this study may be applied to recent advances in optomechanical nanoscopy and spectroscopy, such as photo-induced force microscopy, photothermal-induced resonance microscopy and peak force infrared microscopy. Our analysis will help the interpretation of nanoscale chemical characterization of heterogeneous materials[12, 38] as well as provide insight into the degree of light-matter coupling in metal/van der Waals materials such as surface plasmon polaritons[3] and surface phonon polaritons[4].

## MATERIALS AND METHODS

**Sample preparation.** The polystyrene film is prepared by spin-coating homopolymer of PS onto silicon substrates, which is purchased from Polymer Source Inc. The PS homopolymer has a molecular weight of $M_n$ = 22.5 kg/mol and $M_w/M_n$ = 1.06, resulting in a film thickness of 60 nm. For the SiNc clusters, the diluted Silicon 2,3-naphthalocyanine bis (trihexylsilyloxide) (SiNc) in the toluene from Sigma Aldrich Inc. is spin-coated onto the glass substrate. It resulted in various size of molecular clusters from a few nm to a few hundred nm.

**PiFM measurements.** A VistaScope microscope from Molecular Vista Inc. is used for the mid-IR experiment, which is coupled to a Laser Tune QCL laser system with a wave number resolution of 0.5 cm$^{-1}$ and a tuning range from 800 to 1800 cm$^{-1}$ from Block Engineering. The laser beam is side-illuminated to the sample with angle of 30 degree by an parabolic mirror whose numerical aperture (NA) is around 0.4. The average illumination power was 10 mW with around 20 μm diameter focal spot. The microscope is operated in tapping mode with NCH-Au 300 kHz noncontact cantilevers from Nanosensors. Typically the fundamental resonance is around 300 kHz and the second resonance is around 1.8 MHz. The 30 ns pulsed beam is modulated by tuning its repetition rate to the sum frequency of cantilever's eigenmodes as $f_m = f_2 - f_1 = 1.5$ MHz.

A Nanonics MultiView 2000 tuning-fork based AFM system is used for the near-IR experiment by coupling to a tunable Ti:Sapphire mode Lock laser (Mira 900 from Coherent Inc.). The laser system can switch the continuous wave to femtosecond pulses (< 200 fs, 76 MHz repetition rate) in the range from 700 nm to 980 nm. The laser beam illuminated the sample in an inverted microscope equipped with a high numerical aperture (NA = 1.25) objective lens. The average illumination power was 200 μW with around 400 nm diameter focal spot for the CW experiment and 800 μW for the fs experiment. The microscope is operated in tapping mode with a commercial gold coated tuning-fork from Nanonics company. Typically the fundamental resonance is around 32 kHz and the second mechanical resonance is around 200 kHz. The laser intensity is modulated by using an accusto-optical-modulator at the frequency $f_m$ which is around 168 kHz ($f_m = f_2 - f_1$).

## ASSOCIATED CONTENT

**Supporting Information**. Modeling of the induced charges and the (image) dipole force in the layered system is described in Section 1. The calculation details of the tip-enhanced thermal expansion for 60 nm polystyrene film on Si substrate lies in the Section 2. The calculation details of the modulated van der Waals and Hookian contact force and their force gradient due to oscillatory thermal expansion can be found in the Section 3. This supporting information are available free of charge via the Internet at http://pubs.acs.org.

## AUTHOR INFORMATION

**Corresponding Author**

*Email: eslee@kriss.re.kr (E. S. L)
*Email: epotma@uci.edu (E. O. P)

**Author Contributions**

J. J generated the initial ideas of this work and wrote this manuscript with E. S. L and E. O. P. S. P, D. N and W. M helps J. J to perform the mid-IR PiFM measurements with instrument support. H. K helps J. J to set up the near-IR PiFM and collect the data. E. O. P and E. S. L act as the corresponding authors for this manuscript. All authors contributed to the scientific discussion and manuscript revisions.

**Notes**
The authors declare no competing financial interest.

## ACKNOWLEDGMENT

The authors thank to Dr. Faezeh T. Ladani for the helpful discussion of the dipole force calculation, thank to Mrs. Eun Sook Lee for the help of the graphical visualization. This work was supported by the Korea Research Fellowship Program through the National Research Foundation of Korea (NRF) funded by the Ministry of Science and ICT (2016H1D3A1938071). E.O.P thanks the National Science Foundation (NSF), Grant CHE-1414466, for support.

Table of Contents

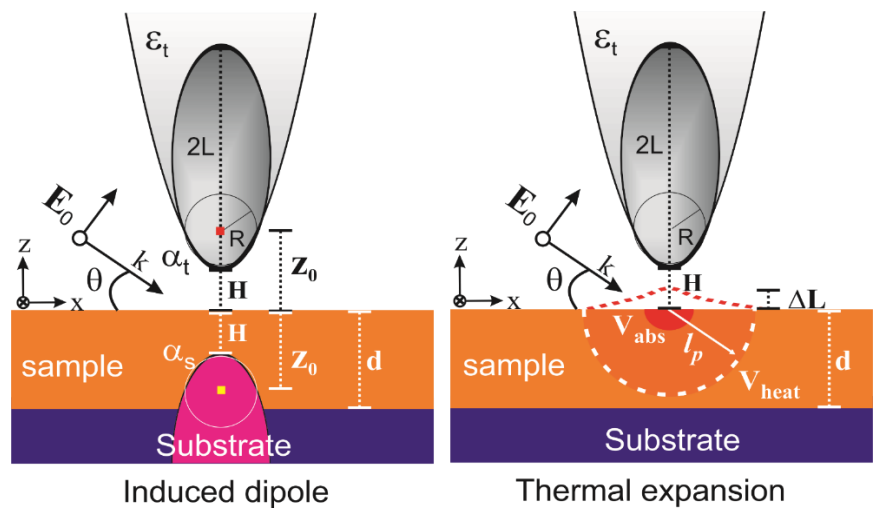

# S1. Derivation of the induced charges and the (image) dipole force in the layered system.

When a sharp metal tip is illuminated, the lightening-rod effect confines fields to the end of the tip. This effect can be modeled with the finite dipole method[1-2]. The schematic diagram in the tip-sample geometry is illustrated in Figure S1a. The tip is modeled as an ellipsoid of length 2L and the tip end is described as a sphere of radius R, which is inscribed in the ellipsoid. The electric field from the tip without the sample substrate is described as[1]:

$$E_s(H) = \frac{\frac{2F(L+H)}{H^2+L(2H+R)}+\ln\frac{L-F+H}{L+F+H}}{\frac{2F(L-\varepsilon_t R)}{LR(\varepsilon_t-1)}+\ln\frac{L-F}{L+F}} E_0 \quad (S1)$$

where F is the focal length given as $F = L\sqrt{1-\frac{R}{L}}$, H is the gap distance from the tip end to the sample surface and the $\varepsilon_t$ is the permittivity of the tip. The electric field can be successfully mimicked by a charge ($Q_0$) which lies in the center of the inscribed sphere. Cvitkovic et al.[1] demonstrated that the s-SNOM tip is well described by adjusting the ratio between the half length of the ellipsoid (L) and the radius of tip (R).

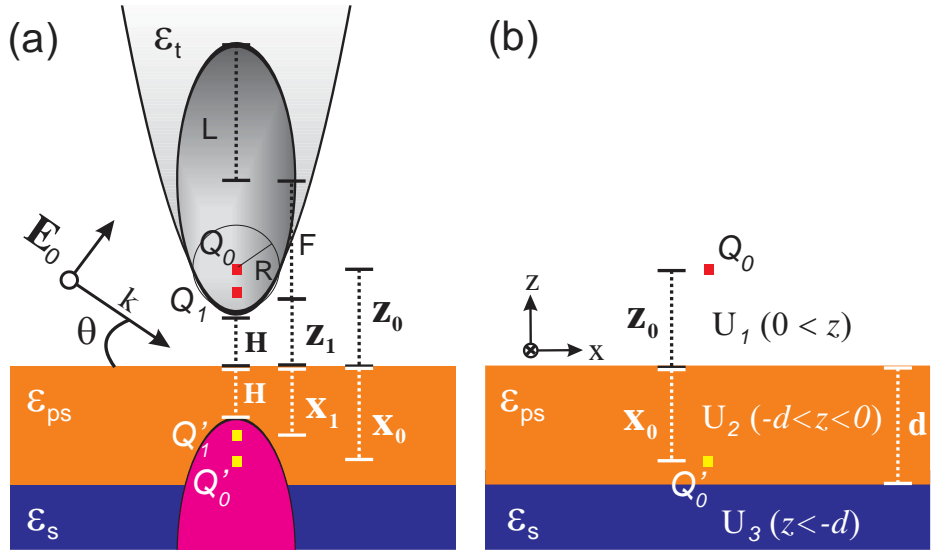

**Fig. S1. Schematic diagram of the image charges with finite dipole method.** (a) Dipole-image dipole interaction based on the ellipsoidal finite dipole model. The image dipole is positioned at $X_0$ in the layered sample. The tip is modeled as an ellipsoid with length of *2L*. The plane wave light is illuminated to the sample with angle of θ. The H is the gap distance from the tip end to the sample surface. (b) Potential responses of a charge $Q_0$ in the distance *z₀* above a flat layered sample.

When the tip is approaching a planar surface, the field enhancement near the tip is boosted by the multiple-scattering process between the tip and the substrate[3]. The process is rigorously described by using the Green function approach. If we focus on the near-field contribution, it can be further simplified by using the image-dipole method. Hauer et al.[2] successfully modeled the multiple-scattering effect in the electrostatic condition by implementing the ellipsoidal finite dipole-image dipole model in the layered system. In this model the two charges are induced on the tip: one is the initial charge ($Q_0$) due to the incident beam $E_0$, which is positioned at the center of the inscribed sphere R. The charge $Q_0$ is described as:

$$Q_0 = 4\pi\varepsilon_0 R^2 E_s(0) = 4\pi\varepsilon_0 R^2 \frac{\frac{2F(L)}{L(a)}+\text{Log}[\frac{L-F}{L+F}]}{\frac{2F(L-\varepsilon_t a)}{La(\varepsilon_t-1)}-\text{Log}[\frac{L-F}{L+F}]} E_0 \quad (S2)$$

where $\varepsilon_0$ is the permittivity of the vacuum. The image charges $Q_0$' due to the external charge $Q_0$ can be described by the potential response U of the sample. The potential response U of the sample to the potential of a charge $Q_0$ at the distance $z_0 = R+H$ above a flat layered sample can be calculated by solving the boundary conditions[4]. The diagram whose origin is located on the sample surface is shown in Figure S1b. The potential above the sample is given as

$$U_1 = -\frac{Q_0}{4\pi\varepsilon_0}(\Phi_0 + \Phi_1),$$

$$\Phi_0 = \int_0^\infty e^{-k|z_0-z|} J_0(kr)dk,$$

$$\Phi_1 = \int_0^\infty A(k) e^{k(z_0-z)} J_0(kr)dk \quad (0<z<z_0). \quad (S3)$$

The potential inside the sample is given as

$$U_2 = -\frac{Q_0}{4\pi\varepsilon_0}(\Phi_2 + \Phi_3),$$

$$\Phi_2 = \int_0^\infty B(k) e^{-k(z_0-z)} J_0(kr)dk,$$

$$\Phi_3 = \int_0^\infty C(k) e^{k(z_0-z)} J_0(kr)dk \quad (-d<z<0). \quad (S4)$$

The potential below the sample (inside substrate) is given as

$$U_3 = -\frac{Q_0}{4\pi\varepsilon_0}\Phi_4,$$

$$\Phi_4 = \int_0^\infty D(k) e^{-k(z_0-z)} J_0(kr)dk, \quad (z<-d) \quad (S5)$$

where $A(k) = \frac{\beta_{12}+\beta_{23}e^{-2kd}}{1-\beta_{21}\beta_{23}e^{-2kd}} e^{-2kz_0}$, $B(k) = \frac{2}{\epsilon_2+1}\frac{1}{1-\beta_{21}\beta_{23}e^{-2kd}}$,

$C(k) = \frac{2}{\epsilon_2+1} \frac{\beta_{23}}{1-\beta_{21}\beta_{23}e^{-2kd}} e^{-2k(d+z_0)}$ and $D(k) = \frac{2}{\epsilon_2+1} \frac{1+\beta_{23}}{1-\beta_{21}\beta_{23}e^{-2kd}}$. For the layered system, the complex electrostatic reflection factor $\beta$ is changed to $\beta_{nm} = \frac{\varepsilon_n - \varepsilon_m}{\varepsilon_n + \varepsilon_m}$ for the layered material where $n$, $m$=1,2,3. According to the boundary conditions, $U_1(z=0) = U_2(z=0)$, $\epsilon_1 \frac{\partial U_1}{\partial z}|_{z=0} = \epsilon_2 \frac{\partial U_2}{\partial z}|_{z=0}$ and $U_2(z=-d) = U_3(z=-d)$, $\epsilon_2 \frac{\partial U_2}{\partial z}|_{z=-d} = \epsilon_3 \frac{\partial U_3}{\partial z}|_{z=-d}$, the potential is continuous at the air-sample surface but the field (normal component) is discontinuous at that boundary. The image charge is described from the potential response $\Phi_1$ because the $\Phi_0$ is the potential response of the external charge $Q_0$ without the sample, while the $\Phi_1$ is the response from the layered sample. The image charge is given as $Q_0' = -\beta_X Q_0$ at the distance $X_0$ under the sample surface (cf. Figure S1b). For the determination of the strength and position of this point charge, we demand that its potential and electric field component in z-direction coincide with the actual response of the sample at $z = 0$:

$$\Phi_1|_{z=0} = -\frac{\beta_X}{z_0+X} \quad \text{and} \quad \frac{\partial \Phi_1}{\partial z}|_{z=0} = \Phi_1'|_{z=0} = -\frac{\beta_X}{(z_0+X)^2}.$$

This condition leads to

$$\beta_X = \frac{\Phi_1^2}{\Phi_1'}|_{z=0} \quad (S6)$$

$$\text{and } X = -\frac{\Phi_1}{\Phi_1'}|_{z=0} \quad (S7).$$

The other induced charge is $Q_1$ due to the mirrored image charges of $Q_0'$ and $Q_1'$. The $Q_1$ is positioned at one of the foci of the ellipsoid, given as $F = L\sqrt{1-\frac{R}{L}}$, which is around half of the radius R/2 from the tip end for R/L = 1/15[1]. The $z_1$ is given as $z_1$=L-F+H≈R/2+H. The $Q_1$ creates an image charge $Q_1'$ which also induces a charge in $Q_1$ through the multiple-scattering process. A self-consistent treatment of the problem leads to the final amount of the induced charge $Q_1$ that has two contributions stemming from the polarization of the sample by $Q_0$ and by $Q_1$ itself[2].

$$Q_1 = \beta_{X_0} f_0 Q_0 + \beta_{X_1} f_1 Q_1 \Rightarrow Q_1 = \frac{\beta_{X_0} f_0}{1-\beta_{X_1} f_1} Q_0 = \eta Q_0 \quad (S8)$$

where $Q_0 = 4\pi R^2 E_s(0)$, $\beta_{X_i} = -\frac{\Phi_1(z_i)^2}{\Phi'_1(z_i)}|_{z=0}$, $f_i = (g - \frac{R+2H+z_i}{2L}) \frac{\ln\frac{4L}{R+4H+2z_i}}{\ln\frac{4L}{R}}$ and $i$=0,1. The g is the empirical geometric factor due to the tip shape. For typical PiFM (or s-SNOM) tip geometries, $|g| = 0.7\pm0.1$[2]. By substituting the $Q_0$->$Q_1$ and $z_0$->$z_1$ into Eq. (S3) to (S7), the potential response of the induced charge $Q_1$ can also be obtained. Together, the total potential response is the sum of the potentials given as $U_{tot}$=U($Q_0$, $z_0$) +U($Q_1$, $z_1$), and the electric field is given by differentiating the potential with respect to the z-axis in the regions.

The electrostatic reflection factor is extended for the layered system as the effective electrostatic reflection coefficient $\beta_{eff}$ as the weighted average of $\beta_{X_0}$ and $\beta_{X_1}$, which is given as[2]:

$$\beta_{eff} = \frac{\beta_{X_0} + \beta_{X_1}\eta}{1+\eta} \quad (S9)$$

where the weighting is done according to the ratio $\eta = \frac{\beta_{X_0} f_0}{1-\beta_{X_1} f_1} = Q_1/Q_0$. The $Re\{\beta_{eff}\}$ decreases from the $Re\{\beta_{Si}\}$ (0.84) to the $Re\{\beta_{PS}\}$ (0.43) as the PS film thickness increases, which plotted in Figure 2b in main text.

The total force between the tip and substrate is generally addressed by using the Maxwell stress tensor[5]. However, this approach requires extensive calculations, which make it more difficult to recognize the physical mechanisms at work. To address this issue, it is possible to implement the dipole approximation using a Green function approach[6]. By focusing on the near-field contribution, this force calculation can be further simplified by the above-mentioned static charge approach. The electrostatic force is directly calculated by using the Coulombic force between the charges on the tip and the sample, which is given as:

$$F_{dip} \approx -\frac{1}{4\pi\varepsilon_0} Re\left\{\frac{\beta_{X_0}|Q_0|^2}{(z_0+X_0)^2} + \frac{\beta_{X_1}|Q_1|^2}{(z_1+X_1)^2} + \frac{(Q_1)^*\beta_{X_0}Q_0}{(z_1+X_0)^2} + \frac{(Q_0)^*\beta_{X_1}Q_1}{(z_0+X_1)^2}\right\}|E_0|^2 \quad (S10).$$

For the spherical point dipole model[7-9] where L=R, Eq. (S10) corresponds to Eq. 1 in main text.

## S2. Tip-enhanced thermal expansion for 60 nm polystyrene film on Si substrate

The electrostatic potential inside (red) and outside (blue) of the 60 nm PS film on Si substrate is plotted in Figure S2a by using Eq. (S3)-(S5) when the gap distance H is fixed as 1 nm. The polystyrene faces with air at z=0 nm and lies in the left side. The tip end is located at z=1 nm by keeping the H=1 nm. $E_1$ (blue) is the gap field between the tip and the sample and $E_2$ (red) is the field is inside the 60 nm polystyrene sample. The normalized electric field distribution is obtained by differentiating the electric potential in Figure S2a with respect to z-axis, plotted in S2b. The simulation parameters are R=30 nm, L=450 nm, $\tau_p$=30 ns, H=1 nm, $\theta$=30 degree and the $\nu_0$ (vibrational resonance)=1452 cm$^{-1}$ which is the C-H bending mode of the PS[10]. The incident electric field should be considered as the peak field for the pulsed beam, which is given as $E_0 = \sqrt{\frac{2}{c\epsilon_0} \frac{I_0}{A} \frac{1}{f\tau_p}}$ where $I_0$ is the incident power, $A$ is the focal area, $f$ is the repetition rate and the $\tau_p$ is the pulse width. Because the radius of our focal area is around 10 um, for $I_0$=10 mW, $A=\pi(10\ um)^2$, $f$=1.6 MHz and $\tau_p$=30 ns, the peak electric field is $E_0$=706785 V/m. By integrating the electric field inside the polystyrene (red solid line) in Figure S2b and substituting it into $P_{abs} = \int a_{abs} \frac{1}{2} c\varepsilon_0 |E|^2 dV$ where $a_{abs} = \frac{4\pi}{\lambda} \frac{9Re[n]Im[n]}{(Re[n]^2+2)^2}$, then one can obtain the tip-enhanced absorption

which increases the temperature inside the sample and eventually results in the strain deformation which gives rise to the thermal expansion. Because we are interested in the maximum thermal expansion to modulate the probe, the maximum temperature change is used for the calculation, descried as:

$$\Delta T_{max} = \frac{P_{abs}}{\rho C V_{heat}} \tau_{rel} \quad (\tau_{rel} < \tau_p) \quad (S11)$$

$$\Delta T_{max} = \frac{P_{abs}}{\rho C V_{heat}} \tau_p \quad (\tau_{rel} > \tau_p) \quad (S12)$$

where $V_{heat}$ is the heated volume related to the thermal wavelength[11], given as $l_p \approx \sqrt{\frac{\kappa_{eff}\tau_p}{\rho C}}$. The tip-enhanced thermal expansion is given as:

$$\Delta L_{max} \approx \sigma d \Delta T_{max} \quad (S13)$$

where $\sigma$ is the linear thermal expansion coefficient. The $\sigma_{ps}$, $\rho_{ps}$, $C_{ps}$, and $\kappa_{ps}$ are 70×10$^{-6}$ K$^{-1}$, 1.05×10$^3$ kg/m$^3$, 1200 J/kg.K and 0.13 W/m.K, respectively, for a PS film. The $\kappa_{eff}$ is the effective thermal conductivity which serially connects the sample material to its surroundings given as $\frac{1}{\kappa_{eff}} \approx \frac{1}{\kappa_{sample}} + \frac{1}{\kappa_{surrounding}} + h$ where the $h$ is the interfacial thermal resistance between the sample and the surrounding. If $\kappa_{surrounding} \gg \kappa_{sample}$, where the polystyrene sample (0.13 W/m.K) is connected to the Si substrate (130 W/m.K), the effective conductivity is reduced to $\kappa_{sample}$ before considering the interfacial thermal resistance. However, for the thin films of low thermal conductivity such as PS is the interfacial thermal resistance between PS and the substrate become the dominant factor for the heat transfer. Unluckily the interfacial thermal resistance is often unknown which make modelling of the thermalization dynamic difficult, but has the general effect of increasing the relaxation time with more prominent effect the thinner the sample. According to Ref. [12] we assume this effect decreases the $k$ of the sample around by 1/2 for the 60 nm PS so that $\kappa_{eff} \approx \frac{1}{2}\kappa_{sample}$ for the low thermal conductivity sample which is connected to the large thermal conductivity substrate. Then the thermal diffusivity, defined as $D \approx \frac{\kappa_{eff}}{\rho C}$, of the polystyrene is 5.16×10$^{-8}$ m$^2$/s and thus the relaxation time is $\tau_{rel} \approx 1.13 \frac{4}{\pi^2} d^2/D \approx 38$ns. Because the relaxation time ($\tau_{rel}$=38 ns) is larger than the pulse duration ($\tau_p$=30 ns), Eq. (S12) is used for the maximum thermal expansion calculation. In this case the maximum thermal expansion is rewritten as:

$$\Delta L_{max} \approx \sigma d \Delta T_{max} \approx \frac{\sigma d P_{abs} \tau_p}{\rho C V_{heat}} \approx \frac{\sigma \tau_p}{\rho C V_{heat}} \frac{A_{abs}}{A_{heat}} \int a_{abs} \frac{1}{2} c \varepsilon_0 |E|^2 \, dz. \quad (S14)$$

We assume the $A_{abs} \approx \pi R^2$ and $A_{heat} \approx \pi l_p^2$ where R=30 nm and $l_p \approx \sqrt{\frac{\kappa_{eff}\tau_p}{\rho C}} \approx 54$nm. By integrating the electric field inside the polystyrene in Figure S3b and substituting it into Eq. (S13), then one can obtain the thermal expansion of ~23 pm at the vibrational resonance of 1452 cm$^{-1}$. The spectral response of the tip-enhanced thermal expansion for the 60 nm PS film on Si substrate is calculated with respect to wavenumber in Figure S2c.

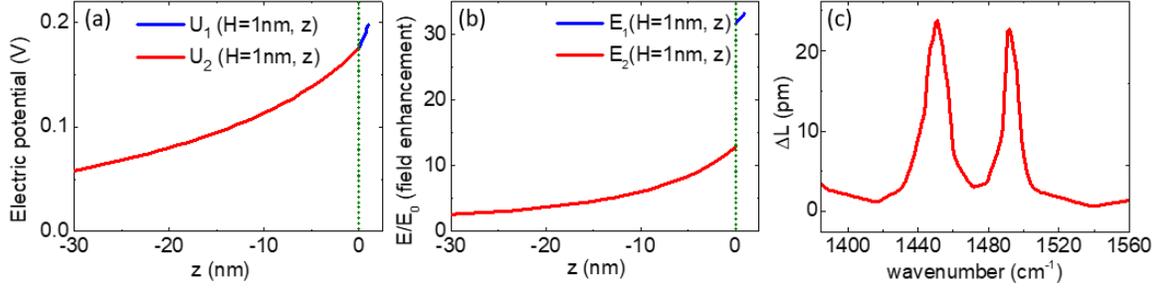

**Figure S2. Thermal expansion based on the tip-enhanced field in the polystyrene film on Si substrate** (a) Calculated electrostatic potential and (b) the normalized field distribution by $E_0$ based on the ellipsoidal finite dipole model. (c) Calculated tip-enhanced thermal expansion for the 60 nm PS film on Si substrate with respect to the wavenumber.

## S3. Calculation of the van der Waals and Hookian contact force and their force gradient due to oscillatory thermal expansion

The modulated the van der Waals and Hookian contact force and their force gradient due to the oscillatory sample expansion are given as below:

$$\Delta F_{vdW} = -\frac{H_{eff}R}{12}\left(\frac{1}{(H-\Delta L)^2} - \frac{1}{H^2}\right) \approx -\frac{H_{eff}R}{6}\frac{1}{H^3}\Delta L(H) \quad (H>r_0) \quad (S15)$$

$$\Delta F_c = k_0(\delta + \Delta L) - k_0\delta \approx k_0\Delta L(r_0) \quad (H<r_0) \quad (S16)$$

$$\Delta \frac{\partial F_{vdW}}{\partial z} = \frac{H_{eff}R}{6}\left(\frac{1}{(H-\Delta L)^3}\left(\frac{\partial H}{\partial z} - \frac{\partial(\Delta L(H))}{\partial z}\right) - \frac{1}{H^3}\frac{\partial H}{\partial z}\right) \approx -\frac{H_{eff}R}{6}\frac{1}{H^3}\frac{\partial(\Delta L)}{\partial H} \quad (H>r_0) \quad (S17)$$

$$\Delta \frac{\partial F_c}{\partial z} \approx \left(k_0\frac{\partial(\delta+\Delta L(r_0))}{\partial z} - k_0\frac{\partial \delta}{\partial z}\right) = 0 \quad (H<r_0) \quad (S18)$$

with $\frac{\partial H}{\partial z} = 1$ and $\frac{\partial \delta}{\partial z} = -1$ where the $H_{eff}$ is the effective Hamaker constant between tip and sample, R is the tip radius, $r_0$ is the interatomic distance (~0.3 nm)[13], $\delta$ is the deflection of cantilever and $k_0$ is the stiffness of the cantilever. The Hamaker constant between polystyrene and the Au tip is described as the reduced Hamaker constant which is given as $H_{eff} = \sqrt{H_{ps} \times H_{Au}}$=1.73 x $10^{-20}$ J where $H_{ps}$ is 6.57 x $10^{-20}$ J and $H_{Au}$ is 4.53 x $10^{-19}$ J. Therefore $\Delta F_{vdW}$ is -640 pN and $\Delta \frac{\partial F_{vdW}}{\partial z}$ is 8.4 N/m for $\Delta L$ = 20 pm, R =

30 nm at H=0.3 nm, and $\Delta F_c$ is 740 pN for $\Delta L = 20$ pm, $k_0 = 37$ N/m and $\Delta \frac{\partial F_c}{\partial z}$ is zero. The tip-enhanced thermal expansion is maximized when the tip is in contact with the surface and it doesn't depend on the deflection of the cantilever (contact force). Because the PiFM sideband signal is proportional to the modulated force gradient, in the dynamic tapping mode PiFM, the modulated van der Waals force gradient is contributed in the PiFM sideband signal in both of the attractive and repulsive region.